\documentclass[aps,prx,reprint,showpacs,amsmath,amssymb,floatfix,superscriptaddress,noeprint]{revtex4-2}
\usepackage{bm}
\usepackage{color}
\usepackage[colorlinks,linkcolor=blue,urlcolor=blue,citecolor=blue,anchorcolor=blue]{hyperref}
\usepackage{graphicx}
\usepackage[capitalise]{cleveref}
\usepackage{physics}
\usepackage[detect-weight=true,separate-uncertainty=true]{siunitx}
\usepackage{multirow}
\usepackage{circledsteps}
\usepackage{comment}

\begin{document}

\title{Amplifying microwave pulses with a single qubit engine fueled by quantum measurements}

\author{R. Dassonneville}
\altaffiliation[Current address: ]{Aix Marseille Université, Université de Toulon, CNRS, IM2NP, Marseille, France}
\affiliation{Ecole Normale Sup\'erieure de Lyon,  CNRS, Laboratoire de Physique, F-69342 Lyon, France}
\author{C. Elouard} 
\altaffiliation[Current address: ]{Université de Lorraine, CNRS, LPCT, F-54000 Nancy, France}
\affiliation{Ecole Normale Sup\'erieure de Lyon, Inria, LIP, F-69342, Lyon Cedex 07, France}
\author{R. Cazali}
\affiliation{Ecole Normale Sup\'erieure de Lyon,  CNRS, Laboratoire de Physique, F-69342 Lyon, France}
\author{R. Assouly}
\affiliation{Ecole Normale Sup\'erieure de Lyon,  CNRS, Laboratoire de Physique, F-69342 Lyon, France}
\author{A. Bienfait}
\affiliation{Ecole Normale Sup\'erieure de Lyon,  CNRS, Laboratoire de Physique, F-69342 Lyon, France}
\author{A. Auffèves}
\affiliation{MajuLab, CNRS-UCA-SU-NUS-NTU International Joint Research Laboratory}
\affiliation{Centre for Quantum Technologies, National University of Singapore, 117543 Singapore, Singapore}
\author{B. Huard}
\affiliation{Ecole Normale Sup\'erieure de Lyon,  CNRS, Laboratoire de Physique, F-69342 Lyon, France}

\date{\today}

\begin{abstract}
Recent progress in manipulating individual quantum systems enables the exploration of  engines exploiting non-classical resources. One of the most appealing is the energy provided by the inherent backaction of quantum measurements. While a handful of experiments have investigated the inner dynamics of engines fueled by measurement backaction, powering a task by such an engine is missing. Here we demonstrate the amplification of microwave signals by an engine fueled by repeated quantum measurements of a superconducting transmon qubit. Using feedback, the engine acts as a quantum Maxwell demon operating without a hot thermal source. Measuring the gain of this amplification constitutes a direct probing of the work output of the engine, in contrast with inferring the work by measuring the qubit state along its evolution. Observing a good agreement between both work estimation methods, our experiment validates the accuracy of the indirect method. We characterize the long-term stability of the engine as well as its robustness to transmon decoherence, loss and drifts. %Our experiment exemplifies a practical usage of the energy brought by quantum measurement backaction.
Our experiment exemplifies the use of energy brought by quantum measurement backaction.
\end{abstract}

\maketitle
%%%%%%%%%%%%%%%%%%%%%%%%%%%%%%%%%%%%%%%
\section{Introduction}

Quantum engines \cite{Myers2022, Cangemi24} are paradigmatic machines designed to explore the thermodynamic impact of  quantum resources - such as coherent superposition or entanglement - as well as of quantum processes triggering energy exchanges. In this context, the inherent backaction of measurements on the state of a quantum system offers a unique tool for manipulating energy flows, specific to the quantum world. When the measured observables do not commute with the system Hamiltonian, the resulting backaction modifies the system average energy, behaving as a non-classical energy resource~\cite{elouard_role_2017,Rogers2022,Mohammady2021,solfanelli_maximal_2019}. This principle allows for the design of quantum engines that have no classical equivalent by incorporating quantum measurement backaction into their thermodynamic cycle. 

Many engines have been proposed in order to exploit measurement-induced energy transfers, where they play a similar role to the heat provided by a hot bath, injecting both energy and entropy into the system, to perform work extraction~\cite{Auffeves2021, Jordan2019}: engines exploiting the coherence of a single qubit requiring a feedback step \cite{brandner_coherence-enhanced_2015, Elouard_2017_PRL, elouard_efficient_2018, WANG2024, Bresque24} or not \cite{Yi2017, Ding2018, su2023thermal}, non-commuting successive measurements \cite{Manikandan2022, Opatrny2021}, continuous monitoring \cite{Manzano2022, Bettmann2023}, general quantum measurements \cite{Behzadi_2021, Purves2021}, or entanglement in two or multi-qubit systems \cite{buffoni_quantum_2019, bresque_two-qubit_2021, Anka2021, Yan2023}. Previous experiments with optical~\cite{Wang2022,Wang23} and NMR setups~\cite{Lisboa2022,Dieguez2023} have demonstrated the feasibility of such engines, for which the work output was inferred from the observed dynamics of the quantum state of the working agent. 
However, a key challenge remains: putting a measurement-fueled engine at work to perform some task on an external physical system. The work provided by the engine can then be directly probed by observing how the external system changes. Unlike the inference of the work from the observation of the working agent dynamics, we argue that this direct monitoring of the work output does not require the engine to be stopped.

Superconducting circuits are highly controllable systems, making them an ideal testbed for quantum energetics and quantum thermodynamics~\cite{Cottet2017, Naghiloo2018, Masuyama2018, Naghiloo2020, Song2021, Pekola2021, Stevens2022, Linpeng2022, Spiecker2023, Gumus2023, Erdamar2024, linpeng2023}. In this study, we implement a quantum engine powered by repeated quantum measurements of a superconducting transmon qubit, following a proposal by some of us~\cite{Elouard_2017_PRL}. Our engine is used to amplify coherent microwave signals propagating in a transmission line (Fig.~\ref{fig:principle}a). Using heterodyne measurements~\cite{Stevens2022}, we determine the gain of this amplification, and thereby directly probe the work extracted by the engine to perform the amplification task. 

The engine operates along the following principle. To fuel the engine, the observable $\hat{\sigma}_x=\ket{-z}\bra{+z}+\ket{+z}\bra{-z}$ is periodically measured. Since this operator does not commute with the qubit Hamiltonian $\hbar\omega_\mathrm{q}\hat{\sigma}_z/2=\hbar\omega_\mathrm{q}(\ket{+z}\bra{+z}-\ket{-z}\bra{-z})/2$,  the measurement backaction gives rise to an input of energy and entropy in the qubit with no classical equivalent~\cite{elouard_role_2017}. Powered up in a superposition of states, the qubit is able to emit a coherent field, which combines with the reflected incoming resonant microwave field. Net amplification requires tuning the phase of the qubit superposition conditionally on the measurement outcome. This feedback step reduces the qubit entropy, at the cost of a decrease of mutual information between the measuring device and the qubit. The experiment can thus be interpreted as a quantum Maxwell demon, where the hot source is replaced by the measurement channel. We also verify that the engine switches off after some time when the measurement is performed but its outcomes are dismissed (open loop configuration), leading to an uncontrolled increase of entropy of the qubit. The switching off is delayed when the repetition rate of the qubit measurement is increased since the measurement outcomes become autonomously more biased by Zeno effect, hence slowing down the entropy growth. Using dispersive measurement readout of the qubit, we perform a full qubit state tomography along its evolution and compare the work we directly probe from the amplification gain to the work that can be inferred from the time-resolved evolution of the qubit state. In contrast with the work of Ref.~\cite{Stevens2022}, no post-selection is involved in the engine, only feedback control. Therefore, the weak values measured in the outgoing power in  Ref.~\cite{Stevens2022} do not contribute to the engine's operation.

The article is structured as follows. In \cref{sec:principle}, we detail the principle of the engine and its physical implementation. Then in \cref{sec:incycle}, we exhibit the correspondence between the measured evolution of the qubit state and the measured gain and extracted work. We also showcase the engine switching off when the Maxwell demon is rendered inactive. Finally, in \cref{sec:W}, we study the performance of gain and work extraction as a function of the amplitude and duration of the driving pulse. 

%%%%%%%%%%%%%%%%%%%%%%%%%%%%%%%%%%%%%%%%%%%%%%%%%%%%%%%%%%%%%%%%%%%%%%%%%%%%%%%%%%%%
\section{Principle of the engine and experimental implementation}
\label{sec:principle}

\begin{figure*}[t!]
    \centering
    \includegraphics{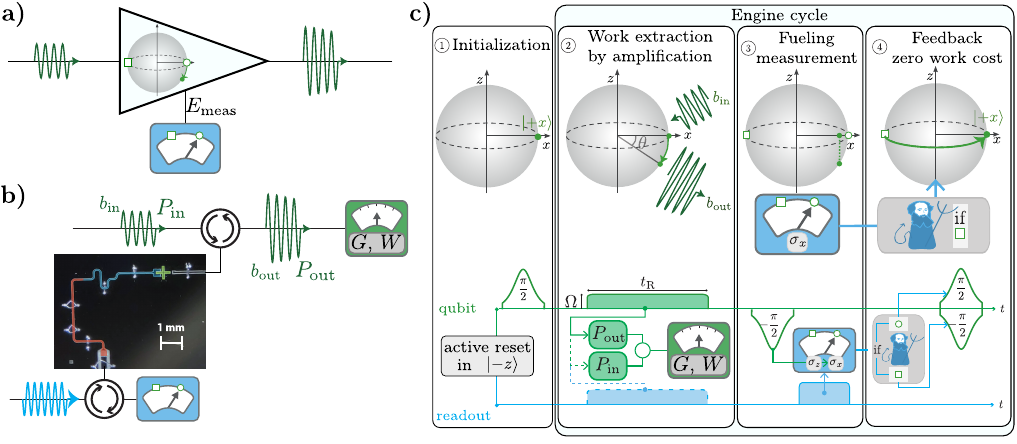}
    \caption{a) A measurement-powered quantum engine amplifying microwave pulses. 
    b) Simplified schematic of the setup with an optical image of the superconducting circuit in false colors. A transmon qubit (green) can be directly driven and its outgoing field is measured via heterodyne detection. The qubit state is read out via its dispersive interaction with a resonator (blue). A Purcell filter (red) inhibits the qubit decay through the readout resonator. 
    c) Principle and pulse sequence of the engine : in \Circled{1}, the qubit is initialized in state $\ket{+x}$ via a $\pi/2$-pulse after a measurement-based active reset in $\ket{-z}$. In \Circled{2}, an incoming resonant microwave signal drives the qubit and is amplified by stimulated emission. Measuring the gain reveals the work output. We measure the output power $P_\mathrm{out}$ during the Rabi drive of amplitude $\Omega$ and duration $t_R$. The useful work is given by $W = (P_\mathrm{out} - P_\mathrm{in}) t_R$ where $P_\mathrm{in}$ (dash lines) is calibrated by detuning the qubit using the AC Stark effect (see \cref{app:calib_P}). The gain is obtained with $G(t) = P_\mathrm{out}/P_\mathrm{in}$.
    In \Circled{3}, the qubit is projectively measured in its $\hat{\sigma}_x$ basis. In \Circled{4}, depending on the outcome of the readout, the qubit is actively reinitialized in $\ket{+x}$ and the thermodynamic cycle restarts at \Circled{2}. The qubit-readout dispersive interaction gives a QND-measurement of $\hat{\sigma}_z$. It is transformed into a QND $\hat{\sigma}_x$ measurement by applying a $-\pi/2$ pulse before and a $+\pi/2$ pulse after the readout pulse. If the qubit is measured in the $\ket{-x}$ state, a $-\pi/2$ pulse is applied instead of the $+\pi/2$ thanks to FPGA-based real-time feedback.}
    \label{fig:principle}
\end{figure*}

After an \textit{initialization} step \Circled{1}, the engine operates on a 3-stroke cycle (\cref{fig:principle}): \Circled{2} \textit{work extraction by amplification of the incoming microwave}, \Circled{3} \textit{quantum measurement} and \Circled{4} \textit{feedback} ~\cite{Elouard_2017_PRL}. We now detail the experimental implementation of each step and the measurement of the gain and work output (see \cref{fig:principle}.b-c and \cref{app:setup,app:calib_P,app:RO,app:dephasing} for more details). 

The initialization step \Circled{1} is achieved by an active reset of the qubit in its ground state $\ket{-z}$, followed by a pulse $+\pi/2$. Active reset is performed by repeatedly applying a $\pi$ pulse to the qubit followed by a readout of $\hat{\sigma}_z$,  until the qubit is found in its ground state. At the end of step \Circled{1}, the qubit thus starts in the state $\ket{+x}=(\ket{+z}+\ket{-z})/\sqrt{2}$ (with a fidelity of \SI{99.2}{\percent}) with an average internal energy $\hbar \omega_\mathrm{q}/2$, where $\omega_\mathrm{q}/2\pi = \SI{4.983}{\giga\hertz}$. 

In step \Circled{2}, the engine performs the amplification task. An incoming square pulse, corresponding to an average Rabi frequency $\Omega$, resonantly drives the qubit for a time $t_\mathrm{R}$. The drive rotates the qubit state by an angle $\theta = \Omega t_\mathrm{R}$ around the $y$-axis of the Bloch sphere, changing its average internal energy by $-\hbar \omega_\mathrm{q} \sin(\theta)/2$. This amount of energy is transferred to the incoming driving pulse, whose coherent amplitude increases via stimulated emission. This energy transfer thus corresponds to an amount of work per cycle $W$ extracted from the qubit to amplify the incoming coherent pulse. From the input/output relation~\cite{Gardiner85,vool_introduction_2017, WallsMilburn,GardinerZoller}, the operator $\hat{b}_{\mathrm{out}}$ associated to the outgoing mode is related to the input mode operator $\hat{b}_{\mathrm{in}}$ and to the qubit lowering operator $\hat{\sigma}_-=|-z\rangle\langle+z|$ via $\hat{b}_\mathrm{out}(t) = \hat{b}_\mathrm{in}(t) + \sqrt{\Gamma_\mathrm{c}} \hat{\sigma}_-(t) $, where $\Gamma_\mathrm{c}/2\pi= \SI{0.383}{kHz}$ is the coupling rate between qubit and the transmission line on which the pulse propagates (\cref{fig:principle}b). The average work flow $P_\mathrm{out}(t)$  outgoing from the qubit at time $t$ is then given by the coherent part of the output power~\cite{cottet_energy_2018,monsel_energetic_2020,Prasad24,MaillettedeBuyWenniger2023Dec}
\begin{align}
    \frac{P_\mathrm{out}(t)}{\hbar \omega_\mathrm{q}} &= \left|\expval{\hat{b}_\mathrm{out}(t)}\right|^2 \notag \\ 
    &=\left|\expval{\hat{b}_\mathrm{in}(t)}\right|^2 + \frac{\Omega}{2}\expval{\hat{\sigma}_x(t)} + \Gamma_\mathrm{c} |\!\expval{\hat{\sigma}_-(t)}\!|^2 \notag \\
    &= \frac{P_\mathrm{in}(t)}{\hbar \omega_\mathrm{q}} + \frac{\Omega}{2}\expval{\hat{\sigma}_x(t)} + \Gamma_\mathrm{c} |\!\expval{\hat{\sigma}_-(t)}\!|^2, \label{eq:expval_pow}
\end{align}
where $\Omega= 2 \sqrt{\Gamma_\mathrm{c}} \left|\expval{\hat{b}_\mathrm{in}}\right|$ is the Rabi frequency and $P_\mathrm{in}(t)$ is the average input power. Moreover, the phase of the input pulse is set such that $\expval{\hat b_\text{in}}$ is real. The term $\Omega\expval{\hat{\sigma}_x(t)}/2$ corresponds to coherent stimulated emission or absorption, while the term $\Gamma_\mathrm{c} |\!\expval{\hat{\sigma}_-(t)}\!|^2$ corresponds to the coherent part of spontaneous emission~\cite{cottet_energy_2018,Prasad24}. 
The gain of the amplification operation performed by the engine is then defined as 
\begin{align}
G(t)&=P_\mathrm{out}(t)/ P_\mathrm{in}(t) \notag \\&= 1 +  \frac{2\Gamma_\mathrm{c} }{\Omega}\expval{\hat{\sigma}_x(t)} +  \frac{4\Gamma_\mathrm{c}^2 }{\Omega^2}  |\!\expval{\hat{\sigma}_-(t)}\!|^2   \label{eq:gain}
\end{align}
The output power also contains an incoherent part associated with the spontaneous emission and given by $P_\mathrm{incoh}/\hbar \omega_\mathrm{q} = \frac{\Gamma_\mathrm{c}}{2} (1+\expval{\sigma_z(t)} - \frac{\vert \expval{\sigma_x(t)} \vert^2}{2} )$ \cite{cottet_energy_2018}.
In the following, we focus on the case where dissipation is much weaker than the drive ($\Gamma_c\ll\Omega$) such that the spontaneous emission is negligible in front of stimulated emission. In this case, we can neglect $P_\mathrm{incoh}$ and also, the last term, in \cref{eq:expval_pow} and in \cref{eq:gain}. 
Heterodyne measurements as performed here provide direct access to the coherent part of the power, which also corresponds to the work flow. Consequently, in the remainder, what we call power has to be understood as its coherent part alone.

Over the duration $t_\mathrm{R}$ of the drive, the average internal energy of the qubit decreases by $\hbar \omega_\mathrm{q} \sin(\theta)/2$, corresponding to the work extracted during the operation. For $\theta \leq \pi$ mod $2\pi$, the extracted work is positive, resulting in a gain $G>1$. It stems from the positive balance between stimulated emission and absorption. To get a sense of the average power levels we need to resolve, let us appreciate that a Rabi rate $\Omega/2\pi$ between $\SI{1}{kHz}$ and $\SI{1}{MHz}$ would produce an excess power $P = P_\mathrm{out}-P_\mathrm{in}$ of at most $\Omega \hbar \omega_\mathrm{q}/2$, which is between $\SI{10}{zW}$ and \SI{10}{aW} for our qubit frequency.

%In step \Circled{3}, the observable $\hat{\sigma}_x$ is measured, yielding the outcome $\pm x$ with probabilities $\cos^2(\theta/2)$ and $\sin^2(\theta/2)$. 
In step \Circled{3}, the observable $\hat{\sigma}_x$ is measured. To do so, we implement a nearly Quantum-Non-Demolition (QND) readout of $\hat{\sigma}_x$ by sequentially performing first a $-\pi/2$ gate on the qubit, followed by a single shot dispersive readout of $\hat{\sigma}_z$, and finally a $\pi/2$ gate (see \cref{app:RO}). The first $-\pi/2$ pulse maps the basis $\{ \ket{-x}, \ket{+x} \}$ onto the $\{ \ket{+z}, \ket{-z} \}$ basis, while the second $\pi/2$ reverses the operation. The overall sequence is exactly equivalent to a projective measurement $\hat{\sigma}_x$. The overall sequence is exactly equivalent to a projective measurement $\hat{\sigma}_x$, yielding the outcome $\pm x$ with probabilities $\cos^2(\theta/2)$ and $\sin^2(\theta/2)$. Whatever the outcome, the measurement deterministically restores the qubit internal energy to its initial value $\hbar \omega_\mathrm{q}/2$. The measurement process, via its backaction on the qubit, is therefore an energy resource for the engine, replacing the hot bath of a conventional heat engine. Implementing the measurement itself however costs resources~\cite{Jacobs09,Guryanova20,Latune25}. A decomposition of the energy flows associated to each pulse in the measurement steps \Circled{3} and \Circled{4} can be found in Ref.~\cite{Elouard_2017_PRL}.
 In the interpretation of the engine as a Maxwell demon, the measurement can be decomposed into two effects. Its backaction provides energy to the qubit and would also increase its entropy if the outcome were discarded. It therefore plays the role of a hot bath. 
Secondly, the information it brings cancels the entropy increase. It corresponds to the demon storing the information into its memory. 
%Here, we experimentally implement a nearly Quantum-Non-Demolition (QND) readout of $\hat{\sigma}_x$ by sequentially performing first a $-\pi/2$ gate on the qubit, followed by a single shot dispersive readout of $\hat{\sigma}_z$, and finally a $\pi/2$ gate (see \cref{app:RO}). The first $-\pi/2$ pulse maps the basis $\{ \ket{-x}, \ket{+x} \}$ onto the $\{ \ket{+z}, \ket{-z} \}$ basis, while the second $\pi/2$ reverses the operation. The overall sequence is exactly equivalent to a projective measurement $\hat{\sigma}_x$.

In the fourth and last step \Circled{4}, the qubit is reset to its initial state $\ket{+x}$ whatever the outcome $\pm x$ obtained in step \Circled{3}. A feedback pulse is applied if and only if the readout outcome is $-x$ (\textit{i.e.}, the Maxwell demon prepares the state for next cycle's work extraction).
As the qubit internal energy is not changed during the feedback step, this Z gate step has no extra energy cost. Experimentally, it is more efficient to merge the Z rotation with the measurement protocol of step \Circled{3} : when the outcome $-x$ is obtained, the $+\pi/2$ gate ending step \Circled{3} is replaced with a $-\pi/2$ mapping $\{ \ket{-z}, \ket{+z} \}$ onto $\{ \ket{-x}, \ket{+x} \}$, while for $+x$,  the $+\pi/2$ is conserved. This feedback operation is performed using Quantum Machines' FPGA-based control system (OPX), and introduces a delay which cannot be reduced below \SI{200}{ns}.

The complete thermodynamic cycle of the measurement engine involves the reset of the memory of the measuring apparatus~\cite{Elouard_2017_PRL}. Here, the memory corresponds to the readout resonator, that leaks out quickly during step \Circled{3} with the characteristic timescale $\kappa^{-1} \sim \SI{18}{ns}$. The steps \Circled{2}-\Circled{4} can therefore be operated cyclically to generate sustained gain and work extraction. 

In contrast with linear amplifiers whose gain is independent on time and signal power up to some saturation point, here the engine amplifies the incoming signal by displacing it by an amplitude and phase set by the qubit coherent superposition (see \cref{app:ampli}). This is a known aspect of stimulated emission, which can be seen as the combination of spontaneous emission with the incoming radiation. Therefore, the gain of the amplification is phase dependent. Further, this displacement is at most $\sqrt{\Gamma_\mathrm{c}}$ and the gain scales as $2\Gamma_\mathrm{c}/\Omega$. It thus decreases with the signal amplitude $\expval{b_\mathrm{in}} \propto \Omega$. Moreover, the amplification has to be triggered and, in general, presents a gain that varies over the course of each cycle. We note that other amplifiers based on single-transmon-qubits in front of a mirror have been proposed and realized \cite{Wen2018,Wiegand2021,Aziz2025}, involving parametric multi-photon processes with a strong pump. When such an amplifier is based on a single two-level qubit, the maximum achievable power gain is \SI{14}{\%} \cite{Wen2018, Wiegand2021}. In contrast, for our engine, the energy required for amplification does not come from a strong pump but from the quantum measurement backaction, and its gain does not present such limitation. 

The direct measurements of the gain and extracted work rely on a calibrated heterodyne measurement, which yields a continuous record of the two quadratures of the outgoing mode $\hat{b}_\mathrm{out}(t)$ (see \cref{app:calib_P}). As the power to be resolved can be as low as tens of zW, the outgoing microwave pulse is amplified using a Traveling Wave Parametric Amplifier (TWPA)~\cite{macklin_nearquantum-limited_2015}. From the heterodyne measurement, we access the instantaneous work flow $P_\mathrm{out}(t)/\hbar \omega_\mathrm{q} = \left\vert \expval{ \hat{b}_\mathrm{out} (t)} \right\vert^2 $ outgoing from the qubit port \cite{Stevens2022} (step \Circled{2} in \cref{fig:principle}.c). The heterodyne detection can also give us access to the instantaneous quadratures of the incoming mode $\hat{b}_\mathrm{in}(t)$ and thus the instantaneous input power $P_\mathrm{in}(t)/\hbar \omega_\mathrm{q} = \left\vert \expval{ \hat{b}_\mathrm{in} (t)} \right\vert^2$ provided we switch off the coupling to the qubit. We do so by shifting the qubit out of resonance with the incoming microwave pulse using the AC Stark shift effect (see \cref{app:dephasing}), thereby suppressing the amplification process. The measured output photon rate $\left\vert \expval{ \hat{b}_\mathrm{out} (t)} \right\vert^2$ then equals $P_\mathrm{in}(t)/\hbar \omega_\mathrm{q}$. The measurements of $P_\mathrm{out}$ and $P_\mathrm{in}$ are interleaved to mitigate as much as possible the detrimental effects of experimental drifts over time. 
The extracted work flow $P(t)$ and the instantaneous gain $G(t)$ are linked via $P(t) = P_\mathrm{out}(t) - P_\mathrm{in}(t)= P_\mathrm{in}(t)(G(t)-1)$.

%%%%%%%%%%%%%%%%%%%%%%%%%%%%%%%%%%%%%%%%%%%%%%%%%%%%%%%%%%%%%%%%%%%%%%%%%%%%%%%%%%%
\section{Characterization of the cyclic evolution, the importance of feedback}
\label{sec:incycle}

\begin{figure}%[h!]
    \centering
    \includegraphics[width=7.5cm]{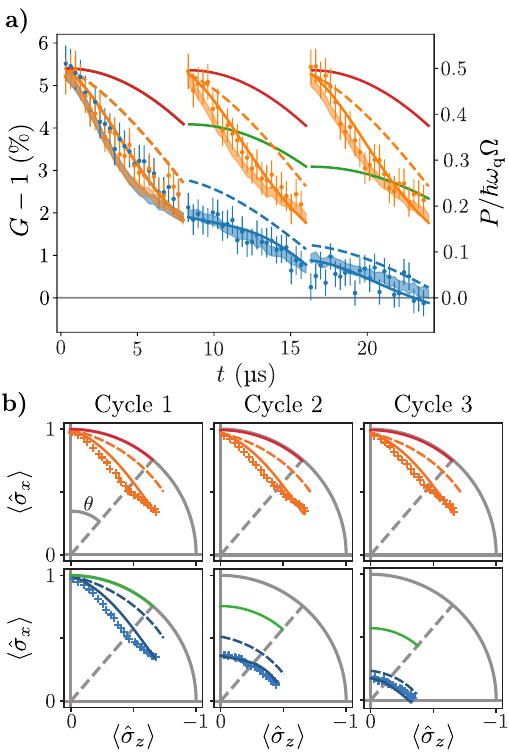}
    \caption{a) Dots: measured excess gain $G-1$ (left) and excess power $P/\hbar \omega_\mathrm{q} \Omega$ (right) as a function of time during the first three cycles of the normal engine (orange) or in open-loop configuration (blue). Here,  $\Omega/2\pi = \SI{14.2}{kHz}$ and $t_\mathrm{R} = \SI{8}{\micro\second}$. Error bars combine the standard deviation divided by the square root of sampling number and a \SI{2}{\percent} relative uncertainty on the calibrated gain $G_\mathrm{meas}$. Shadows: inferred excess gain and power using the tomography of the qubit state in b) with Eqs.~(\ref{eq:expval_pow}),(\ref{eq:gain}), and assuming a \SI{1}{\percent} uncertainty due to gate and readout infidelities. Solid red (green) lines: case of infinite $T_1$ and $T_2$  and resonant drive of the (open-loop) engine. Dashed lines: result of numerical simulations assuming resonant drive, $T_1=\SI{25}{\mu s}$ and $T_2=\SI{32}{\mu s}$. Solid orange/blue lines: result of simulation assuming a distribution of qubit frequencies and coherence times $T_2$ (see \cref{app:stability} and \cref{app:simulations}). The visible dead times of \SI{536}{ns} correspond to the duration $ t_\mathrm{meas}$ of steps 3 and 4 in the engine cycle. b) Crosses: measured Bloch coordinates $\expval{\hat{\sigma}_x(t)}$ as a function of $\expval{\hat{\sigma}_z(t)}$ during the same first three cycles of the engine from left to right (top, orange) or in the open-loop configuration without feedback (bottom, blue). Other lines correspond to their counterpart in a).
    } 
    \label{fig:cyclic_evol}
\end{figure}

We start by running the engine and measure the gain $G(t)$ and work flow $P(t)$  within the first three cycles after initialization. This measurement is shown as orange dots in \cref{fig:cyclic_evol}a for a work extraction duration $t_R = \SI{8}{\mu s}$, and a Rabi frequency $\Omega/2\pi = \SI{14.2}{kHz}$. It is averaged over hundreds of thousands of repetitions for $G(t)$ and $P(t)$. The measurement clearly demonstrates that the engine amplifies the incoming signal by a few \%. Furthermore, the measured instantaneous excess gain and power are periodic, which shows that the engine operates in a cyclic manner.

It is possible to compare the measured excess power to the extracted work that can be inferred from the qubit tomography. Note that, in contrast to direct probing, tomographic probing requires stopping the engine in order to measure a qubit observable at an intermediate time within the cycle. Indeed, the measurement backaction associated with the qubit state tomography alters the qubit subsequent evolution. By repeating the process for a complete set of qubit observables, and by varying the measurement time, one can reconstruct the state evolution from which gain and work are inferred. We observe that the qubit follows the same trajectory in the Bloch sphere for all three cycles  (orange crosses in \cref{fig:cyclic_evol}b). The visible change in $\hbar\omega_\mathrm{q}\langle\hat{\sigma}_z\rangle/2$ between the end of one cycle and the beginning of the next corresponds to the energy that was provided by the measurement backaction in steps 3 and 4. 
We represent the expected excess gain, power and qubit state evolution in the case of infinite qubit coherence time as red solid lines in \cref{fig:cyclic_evol}. However, due to the relatively long drive duration $t_R$ compared to the finite coherence times $T_1$ and $T_2$, the qubit does not remain pure (on the surface of the Bloch sphere) within a cycle and thus is not fully captured by the Rabi angle $\theta$. Moreover, we note that the measured qubit state evolution during step \Circled{2} cannot be reproduced by taking only into account the finite values of $T_1$ and $T_2$ (dashed lines in \cref{fig:cyclic_evol}b). A model taking into account the spurious fluctuations of the qubit frequency and qubit coherence times (see \cref{app:stability}), improves the agreement between simulations and qubit state evolution, excess gain and power (solid lines in  \cref{fig:cyclic_evol}a-b).

Plugging this qubit state evolution into Eqs.~(\ref{eq:expval_pow}),(\ref{eq:gain}), we obtain the inferred excess gain and work flow (shaded regions in \cref{fig:cyclic_evol}a. We find a good agreement between these quantities and the measured excess gain (G-1) and power $P(t)$. This agreement shows that the microwave mode has received nearly as much work as the qubit has provided.
We expect the small deviations between the two to mostly originate from the slow drift of qubit parameters over time. Indeed, qubit state evolution and direct probing measurements were not taken simultaneously and as such were prone to different realizations of the fluctuating qubit parameters. As the qubit state measurements take less averaging and thus less time than the direct measurement of the gain and power, they are less prone to drifts. 

The model with fixed qubit parameters (dashed lines in \cref{fig:cyclic_evol}a-b) overestimates the measured gain $G(t)$ and power $P(t)$, as well as the qubit state purity. Indeed, parasitic changes of qubit frequency over time are detrimental to the engine's operation since a resonant Rabi drive is assumed and required. The resulting detuning $\Delta = \omega_\mathrm{d}-\omega_\mathrm{q}$ of the drive does not only increase the Rabi rate as $\Omega' = \sqrt{\Omega^2 + \Delta^2}$ but also shifts the rotation axis. A more complex trajectory on the Bloch sphere is then expected if there is an uncontrolled detuning $\Delta$ during the Rabi drive duration $t_R$, which might even result in negative extracted work. We simulate the gain and extracted work by assuming that the qubit takes only a few discrete values of $\omega_\mathrm{q}$ and $T_2$ due to its coupling to uncontrolled two-level-systems (TLS)~\cite{Muller_2019}. To estimate this effect, we have measured the qubit parameters $\omega_\mathrm{q}$, $T_1$ and $T_2$ as a function of time with relaxation and Ramsey measurements in a separate experimental run (see~\cref{app:stability}). In a more complete model, we take into account the role of two main TLS only, one shifting the qubit frequency by \SI{70}{kHz} with $T_1=\SI{25.4}{\micro s}$ and $T_2 = \SI{34}{\micro s}$, and the other one shifting the qubit frequency by \SI{6}{kHz} but also reducing the coherence to $T_2 = \SI{21}{\mu s}$. We assume that, over the many samples averaged for a given measurement, each TLS has a probability $p$ to be in its excited state. The only fitting parameters in our model are these probabilities (see~\cref{app:stability}). The resulting predicted gain and power are plotted as a solid orange line in \cref{fig:cyclic_evol}a. Less significant contributions to the discrepancy between shadow areas (inferred from qubit tomography) and the dots (direct measurement) can be attributed to the last term in Eq.~(\ref{eq:expval_pow}) and to the entanglement between qubit and outgoing field  \cite{Stevens2022}.

For comparison, we also performed the qubit tomography and engine gain and power measurement in the case of the open-loop configuration by purposely dismissing the feedback step \Circled{4}. It is realized by always performing a $+\pi/2$ pulse whatever the readout outcome with the same delays as when the feedback is on. In that case, the purity of the qubit state decreases over the cycle, such that the cycles are not equivalent to each other (blue crosses in \cref{fig:cyclic_evol}b). Assuming an ideal qubit with no decoherence, as the randomness of the measurement-induced dynamics effectively produces the mixed qubit state $\rho^{(N_c)}=p_{+}^{(N_c)}\ket{+x}\bra{+x}+ p_{-}^{(N_c)}\ket{-x}\bra{-x}$ at the end of the $N_c$-th cycle, with $p_{\pm}^{(N_c)}= [1 \pm \cos^{N_c}(\theta) ]/2 $, the qubit state gets closer to the center of the Bloch sphere after each cycle (green lines in \cref{fig:cyclic_evol}b). The excess power $P^{(N_c)} = \frac{\hbar \omega_\mathrm{q} \Omega}{2} \cos^{N_c}(\theta)$ and excess gain $G^{(N_c)} -1 = \frac{2\Gamma_\mathrm{c}}{\Omega} \cos^{N_c}(\theta)$ thus tend to zero with the number of cycles (measured excess gain and power as blue dots in \cref{fig:cyclic_evol}a). The same simulations as for the normal engine operation can be performed in this open-loop case (blue lines in \cref{fig:cyclic_evol}). They reproduce the measurements similarly well. 

In the open-loop configuration, power could still be extracted for numerous cycles in the small angle limit $\theta \rightarrow 0$, where the excess power at cycle $N_c$ is approximately given by $ P^{(N_c)} \simeq \frac{\hbar \omega_\mathrm{q} \Omega}{2} (1 - \frac{N_c \theta^2}{2})$. This corresponds to the Zeno limit where a strong $\hat{\sigma}_x$ measurement ``freezes" the qubit in its $\ket{+x}$ state, making measurement back-action deterministic, and the feedback of the Maxwell demon unnecessary \cite{Elouard_2017_PRL}. However, a clear demonstration of this regime would be difficult owing to the limited coherence time $T_2$ of the qubit, which is not large enough compared to the duration of the measurement of $\hat{\sigma}_x$ (\cref{app:feedbackoff}). Interestingly, due to the finite coherence times, the open-loop configuration can present a negative extracted work in the steady state, especially when $t_R$ is a non-negligible fraction of $T_1$ and $T_2$. For instance, in the case where $t_R = \SI{8}{\mu s}$ and $\Omega/2\pi = \SI{20.2}{kHz}$, we observe, in the steady state, a negative excess power of about $\SI{-0.02}{aW}$, which is absorbed on average by the qubit which ends up closer to $\ket{-x}$ than $\ket{+x}$ at the end of each cycle (see Fig.~\ref{fig:P_vs_Nc} in \cref{app:PvsNc}).  Finally, we note that applying the feedback loop only intermittently has been recently shown to be rewarding in terms of performance \cite{Bresque24}.

\section{Gain and extracted work as a function of signal amplitude and cycle duration}
\label{sec:W}

\begin{figure}[h!]
    \centering
    \includegraphics[width=8.6cm]{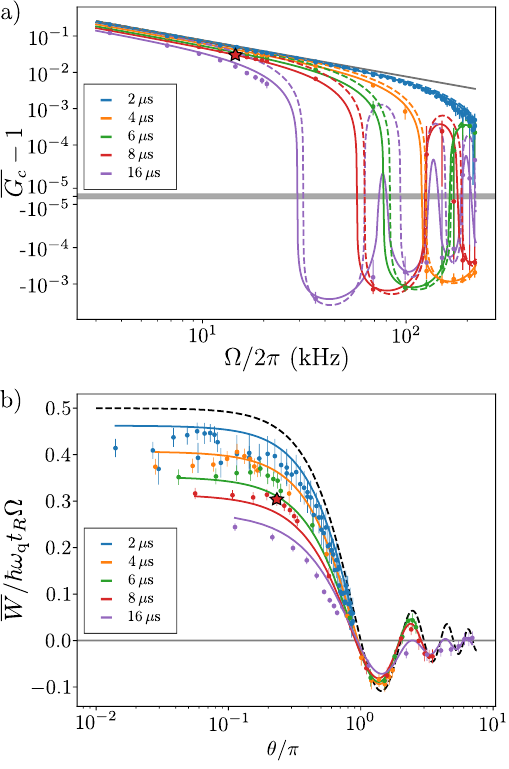}
    \caption{
    a) Dots: the mean measured excess gain $\overline{G_c}-1$ averaged over 40 cycles as a function of Rabi frequency $\Omega$ in log-scale representation. Note that the log scale is cut around zero (gray area). Colored dashed lines : prediction  for an ideal qubit with infinite lifetime and coherence times and resonant driving. Solid gray line : small angle limit $2\Gamma_\mathrm{c}/\Omega$. b) Dots: normalized extracted work per cycle  $\overline{W}/\hbar\omega_\mathrm{q}t_R\Omega$ averaged over 40 cycles as a function of Rabi angle $\theta = \Omega t_\mathrm{R}$. Black dashed line: prediction for an ideal qubit with infinite lifetime and coherence times and resonant driving.
    a-b) Error bars combine the standard deviation divided by the square root of sampling number and a \SI{2}{\percent} relative uncertainty on the calibrated gain $G_\mathrm{meas}$. The different colors correspond to different durations $t_\mathrm{R}$ (see legend). Solid lines are numerical simulations assuming a distribution of qubit frequencies and coherence times $T_2$ (see \cref{app:stability} and \cref{app:simulations}). The star indicates the parameters used in \cref{fig:cyclic_evol}.}
    \label{fig:work_theta}
\end{figure}

Integrating the instantaneous gain $G(t)$ and power $P(t)$ over the drive duration $t_R$ within each cycle, we obtain the cycle gain $G_c(N_c) = \frac{1}{t_R}\int_{(N_c-1)t_\mathrm{cycle}}^{(N_c-1)t_\mathrm{cycle}+t_R} G(t) dt$, with $t_\mathrm{cycle}=t_R+t_\mathrm{meas}$, and extracted work over one cycle $W(N_c) =  \int_{(N_c-1)t_\mathrm{cycle}}^{(N_c-1)t_\mathrm{cycle}+t_R}  P(t) \mathrm{d}t$, which we observe to be, as expected, a constant of the cycle number $N_c$ (see \cref{app:PvsNc}). We further study the performance of the engine (feedback on in step \Circled{4}) by plotting the mean excess gain $\overline{G_c}-1$ (dots in \cref{fig:work_theta}a)  
averaged over 40 cycles as a function of $\Omega$ for values of $t_R$ in [2, 4, 6, 8, 16] \si{\micro s}. For each duration $t_R$ of the work extraction step, the excess mean gain oscillates as a function of incoming drive amplitude $\left|\expval{\hat{b}_\mathrm{in}}\right|=\frac{\Omega}{2 \sqrt{\Gamma_\mathrm{c}}}$. For low amplitudes, the qubit Bloch vector rotates by less than $\pi$, which leads to net extracted work and hence $\overline{G_c}>1$. However, when the amplitude is large enough, the qubit rotates by more than $\pi$, which costs work overall, hence de-amplifying the driving signal ($\overline{G_c}<1$). This transition can be better observed in the mean normalized extracted work $\overline{W}/\hbar \omega_\mathrm{q}t_R \Omega$ (dots in \cref{fig:work_theta}b) averaged over 40 cycles as a function of $\theta = \Omega t_R$ for the same values of $t_R$. Due to the finite qubit coherence time $T_2$ and the fast changes and drifts of the qubit parameters, the rotation angles $\theta$ at which the mean work changes sign are not exactly multiples of $\pi$. 

The behavior of the measured mean gain and work is qualitatively reproduced by the simple limit of infinite qubit lifetime and coherence time. In this ideal case, the excess gain is given by $\overline{G_c}-1 = 2\Gamma_\mathrm{c}/\Omega ~ \mathrm{sinc}(\theta)$ while the normalized extracted work is $\overline{W}/\hbar \omega_\mathrm{q}t_R \Omega = \mathrm{sinc}(\theta)/2$~\cite{Elouard_2017_PRL}. 
This limit case is represented by colored dashed lines in \cref{fig:work_theta}a, which translates into the single black dashed line in \cref{fig:work_theta}b. 

For small angles $\theta$, the mean gain and extracted work are smaller for longer work extraction times $t_R$. 
As the driving duration $t_R$ becomes a non-negligible fraction of $T_2$, the entropy of the qubit increases within each cycle and the excess gain and of extracted work are reduced since $\overline{G_c}-1 = \frac{2\Gamma_\mathrm{c}}{\Omega} \frac{1}{t_R}\int_0^{t_R} \cos(\Omega t) e^{-t/T_2} dt$ and $\overline{W}/\hbar \omega_\mathrm{q} \Omega = \frac{1}{2} \int_0^{t_R} \cos(\Omega t) e^{-t/T_2} dt$. These expressions with fixed coherence time do not fully capture the observed excess gain and extracted work. As in the previous section, we have to consider the fast changes and drifts of the qubit parameters in order to better reproduce the measurements, leading to the solid lines in Fig.~\ref{fig:work_theta}a-b (see also section \cref{app:stability}). Under these assumptions, the excess gain $\overline{G_c}-1$ and extracted work $\overline{W}$ as a function of $\Omega$ and $t_R$ are well reproduced.

In the small angle limit $\theta \ll 1$, the qubit stays close to the state $\ket{+x}$ where $\expval{\hat{\sigma}_x} = 1$, so that the maximal mean work extraction and excess gain are respectively predicted to be $\hbar\omega_\mathrm{q} t_R \Omega/2$ and $2\Gamma_\mathrm{c}/\Omega$ (solid gray line in \cref{fig:work_theta}a).
Reaching this maximum requires fulfilling the condition $T_2 \gg \Omega^{-1} \gg t_R \gg t_\mathrm{meas}$, resulting in, for our device, $ \SI{30}{\mu s} \gg \Omega^{-1} \gg t_R \gg \SI{536}{ns}$. 
 The measured excess gain and extracted work can be seen to approach these upper bounds for the smallest values of $t_R$ and $\Omega$ or $\theta$. For our lowest value of $\Omega$, we observe an excess gain ranging from \SI{13}{\percent} to \SI{21}{\percent} for $t_R$ from $\SI{16}{\mu s}$ to $\SI{2}{\mu s}$ respectively. It is a bit below the theoretical higher bound $2\Gamma_\mathrm{c}/\Omega$ achieved at small angles, which is here about $\SI{25}{\percent}$ for $\Omega/2\pi=\SI{3.3}{kHz}$. We note that the lowest value of $\Omega$ exits from the regime $\Gamma_\mathrm{c} \ll \Omega$ where the incoherent part of the power is negligible. More precisely, the ratio ($\Gamma_\mathrm{c}/\Omega$ reaches about \SI{12}{\percent}), leading to a fraction $\Gamma_\mathrm{c}/2\Omega \simeq \SI{6}{\percent}$ of the energy brought by the measurement to be dissipated as heat via spontaneous emission rather than extracted as work.

We finish our analysis by discussing the efficiency of the engine. The thermodynamic task it performs during each cycle is to convert the amount of energy $E_\text{meas}$, provided by the measurement setup, into extracted work $W$ (\cref{fig:principle}a). In analogy with the analysis of a conventional Maxwell demon engine converting the heat from a hot bath into work, an efficiency can be associated to this conversion process via $\eta = (W-W_{er})/E_\text{meas}$, where $W_{er}$ stands for the erasure work and is proportional to the final entropy of the memory of the measuring device. This quantity was analyzed in Ref.~\cite{Elouard_2017_PRL} and showed theoretically to reach unity in the limit $\theta\to 0$, where the power is also maximized. In the present experiment, the energy $E_\text{meas}$ is provided by the microwave pulses forming the measurement sequence steps \Circled{3} and \Circled{4}. An estimate of $E_\mathrm{meas}$ would then consist in computing the energy of the pulses involved in the $\sigma_x$-measurement in steps \Circled{3} and \Circled{4}. In our implementation, the microwave power arriving at the sample are about \SI{0.1}{pW} for the $\pm \pi/2$ pulses and about \SI{4}{fW} for the readout pulse, meaning in total an energy of about $\sim$1600 energy quanta at the qubit frequency.
Further tracking the thermodynamic nature of the energy $E_\text{meas}$ in the measurement apparatus, and comparing it to the total cost to perform the measurement is beyond the scope of this experiment. It relates to the question of building a consistent thermodynamic model of the measuring apparatus, to predict the energy transfers accompanying the measurement process and its fundamental costs, which is an active field of research \cite{Jacobs09,Deffner2016,Mancino2018,Guryanova20,bresque_two-qubit_2021,Piccione-PRL2024,Linpeng-PRL-23, linpeng2023,Mohammady23,Latune25}.

\section{Conclusion}
We have successfully demonstrated the operation of an engine solely powered by quantum measurement backaction and put to the task of amplifying an incoming microwave signal. %The engine is able to increase the incoming power by more than $10~\%$. 
Probing the gain of this amplification provides a direct measurement of the work extracted by the engine. We have linked the power flow inferred from qubit tomography at various times to the continuously measured excess power in the outgoing microwave signal within a cycle and as a function of cycle number. 
We have identified two primary limitations to the performance of the quantum measurement engine: first, the frequency stability of the qubit over time, and second, its finite coherence time. Such limitations may be mitigated with better optimization of the superconducting qubit fabrication. Moreover, the $\hat{\sigma}_x$ readout outcomes over several successive cycles may be used to estimate the qubit detuning and stabilize the qubit frequency during the engine operation with ac-Stark effect or using a flux-tunable qubit \cite{Vepsalainen2022}.

Our current focus has been on measuring the average gain and extracted work. However, in future studies, investigating the full statistics of work fluctuations through single-shot measurements would be of great interest, especially in the context of quantum fluctuation theorems \cite{Elouard19,Prech24} and thermodynamic/kinetic uncertainty relations \cite{Hasegawa20,Reiche22,Maity24,Moreira24}. Achieving this would require work measurement with high quantum efficiency and high resolution, such as employing high-efficiency heterodyne detection \cite{Eddins2019}, 
microwave photon detectors or counters~\cite{dassonneville_number-resolved_2020, Lescanne2020, Balembois2024} or bolometers and calorimeters \cite{gunyho2024zeptojoulecalorimetry}. The reflection of the relatively large incoming drive may be challenging to handle. This issue can be mitigated by using non-resonant gates and/or Rabi drives such as subharmonic driving \cite{xia2023} in order to separate the incoming field from the outgoing field in frequency.  
 
Superconducting quantum circuits, enabling high-fidelity QND measurements and feedback schemes, present a state-of-the-art platform for investigating the thermodynamic costs of quantum information processing, particularly in studying the performance and limitations of quantum measurement engines with different cycling strategies ~\cite{brandner_coherence-enhanced_2015, elouard_efficient_2018, WANG2024, Yi2017, Ding2018, su2023thermal, Manikandan2022, Opatrny2021, Manzano2022, Bettmann2023, Behzadi_2021, Purves2021, buffoni_quantum_2019, bresque_two-qubit_2021, Anka2021, Yan2023, Auffeves2021, Elouard_2017_PRL}. Advancing the study of quantum measurement engines will not only improve our understanding of the thermodynamic costs and energy efficiency of quantum information processors~\cite{Abah_2019, Gea-Banacloche2002,ikonen_energy-efficient_2017, Silva_Pratapsi2023, gois2024energetic, Fellous-Asiani2023}, but will also allow exploring the foundations of quantum measurement with energetic probes~\cite{Jordan2019,Piccione-PRL2024,Revue-Nature25}.

\begin{acknowledgments}
We acknowledge IARPA and Lincoln Labs for providing a Josephson Traveling-Wave Parametric Amplifier. 
A.A. acknowledges the National Research Foundation, Singapore through the National Quantum Office, hosted in A*STAR, under its Centre for Quantum Technologies Funding Initiative (S24Q2d0009), the ANR Research Collaborative Project “Qu-DICE” (Grant No. ANR-PRC-CES47), the Foundational Questions Institute Fund (Grant No. FQXi-IAF19-01 and Grant No. FQXi-IAF19-05) and the John Templeton Foundation (Grant No. 61835). We thank Guillaume Cauquil, Maria Maffei, and Antoine Marquet for useful discussions. 
\end{acknowledgments}

%%%%%%%%%%%%%%%%%%%%%%%%%%%%%%%%%%%%%%
\appendix
%%%%%%%%%%%%%%%%%%%%%%%%%%%%%%%%%%%%%%
\section{Sample and measurement setup}

\label{app:setup}

\begin{figure}[h!]
    \centering
    \includegraphics[width=8.6cm]{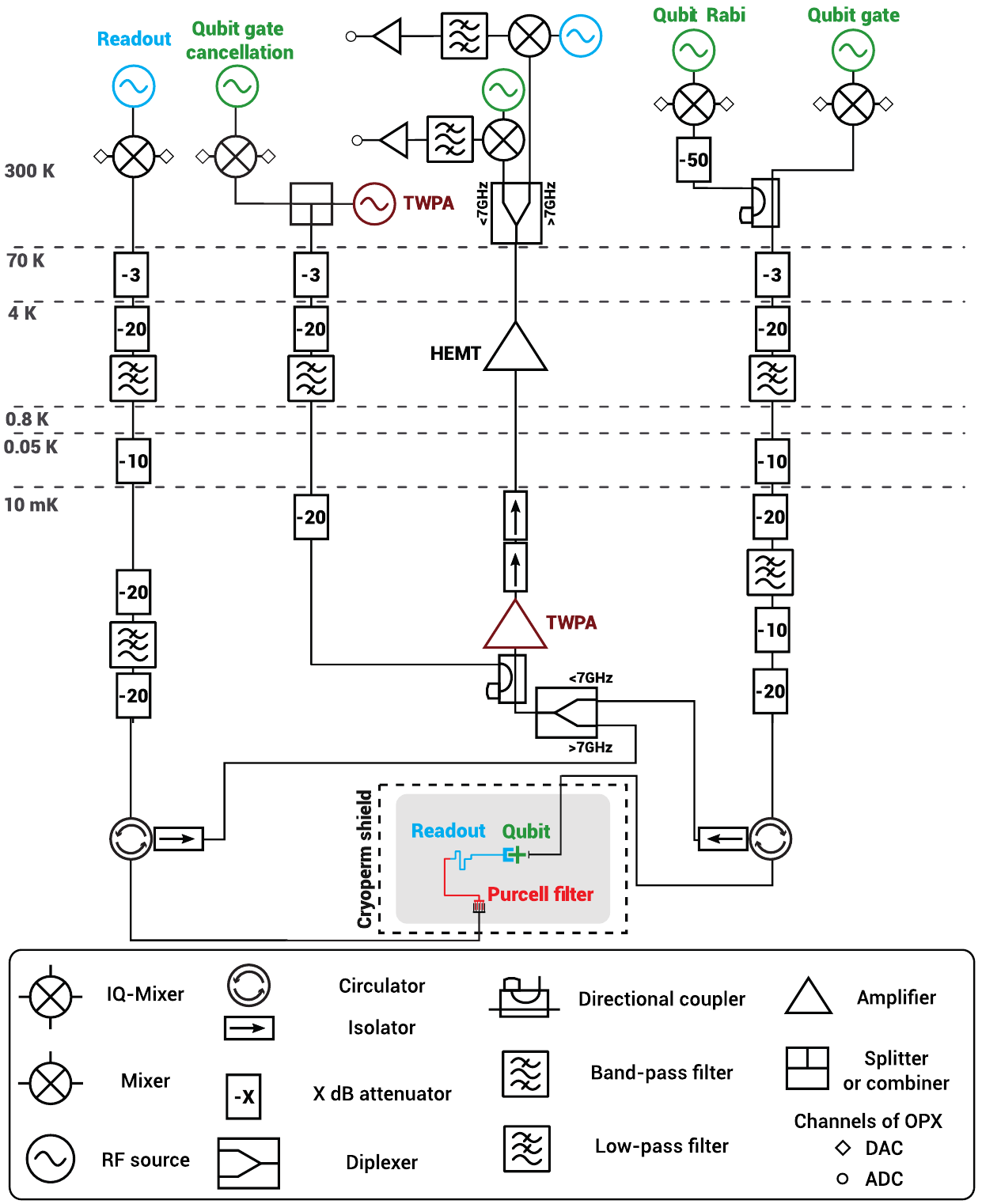}
    \caption{Schematic of the microwave wiring. The Josephson Traveling Wave Parametric Amplifier (TWPA) was provided by the Lincoln Lab. The RF source colors refer to the frequency of the matching element in the device up to a modulation frequency. Identically colored sources represent a single instrument with a split output.}
    \label{fig:wiring}
\end{figure}

The large features of the circuit are made by optical lithography on a Tantalum thin film deposited on a Sapphire substrate. The readout mode is a $\lambda/4$ coplanar waveguide resonator. The Purcell filter is also a $\lambda/4$ coplanar waveguide resonator, with frequency close to and inductively coupled to the readout mode. It is used as a bandpass filter around the readout frequency and limits the qubit decay through the readout resonator line. The Josephson junction of the transmon qubit is fabricated using electronic lithography followed by angle deposition of Al/AlOx/Al in a Plassys evaporator.

The sample is cooled down to \SI{10}{mK} in a dilution refrigerator. The diagram of the microwave wiring is given in \cref{fig:wiring}. The microwave pulses are generated by modulation of continuous microwave tones produced respectively by generators Anapico APSIN20G and Anapico APSIN12G (all qubit-type tones use the same local oscillator). They are modulated via IQ-mixers where the intermediate frequency (a few tens of \si{MHz}) modulation pulses are generated by 8 DAC channels of an OPX from Quantum Machines with a sample rate of \SI{1}{GS\per \second}. Two channels are used for the readout operation, while the 6 others used for the qubit operation are separated into 3 sets of 2, namely, qubit Rabi, qubit gate and qubit gate cancellation. The qubit Rabi and qubit gate pulses are sent into a single transmission line using a directional coupler. The qubit gate is used to generate the fast $\pi$ and $\pi/2$ qubit gate pulses while the qubit Rabi is used to generate the longer drive at amplitude $\Omega$ and duration $t_R$ with a square envelope. Due to the large dynamical range (around \SI{70}{dB} difference) required between qubit gate and qubit Rabi pulses, we used two different IQ-mixers and 4 DAC channels instead of just one IQ-mixer and 2 DAC channels. Moreover, another set of two DAC channels has to be added in order to reduce nonlinear effects such as saturation in the output amplification chain (see below).

The readout and qubit Rabi outgoing pulses are multiplexed into a single output transmission line using a diplexer at the lowest temperature stage. The readout and qubit signals are then amplified with at first a quantum-limited parametric amplifier (a Travelling-Wave Parametric Amplifier, TWPA from Lincoln Lab) and then by a HEMT amplifier. At room temperature, the readout and qubit Rabi signals are separated using a diplexer. Each signal is acquired, after down-conversion by its local oscillator, by digitizing at \SI{1}{GS\per \second} with one of the two ADCs of the OPX. 

In principle, the qubit gate pulses could also be acquired through the same ADC as the qubit Rabi signal. Qubit gate and qubit Rabi signals could be distinguished as they are separated in time during the pulse sequence. However, due to the large power difference between the gate and Rabi pulses, the Rabi detection was impaired \cite{boselli2025}. In order to not disturb the amplification chain with the high-power gate pulses, we have added a gate cancellation tone at the input of the TWPA (see \cref{fig:wiring}). This tone is calibrated to destructively interfere with the reflected qubit gate $\pi/2$ pulses. We then recalibrate our gate pulses with the corresponding applied cancellation gate pulses, as the latter leak towards the qubit with an attenuated power by $\sim \SI{60}{dB}$. 

The transmon qubit has a  frequency $\omega_\mathrm{q}/2\pi = \SI{4.983}{\giga\hertz}$ and anharmonicity of $\SI{-221}{MHz}$. It is coupled to a $\lambda/4$ coplanar waveguide (CPW) readout resonator of frequency $\omega_\mathrm{r}/2\pi = \SI{7.665}{\giga\hertz}$, resulting in the dispersive interaction $\hat{H}/\hbar = -\chi/2 \ \hat{r}^\dag \hat{r} \hat{\sigma}_\mathrm{z}$ with coupling rate $\chi/2\pi = \SI{3.3}{MHz}$ where $\hat{r}$ is the annihilation  operator of the readout resonator. The readout resonator is also coupled to a $\lambda/4$ CPW bandpass Purcell filter, resulting in a readout decay rate $\kappa/2\pi = \SI{9}{MHz}$. The Purcell filter is intended to limit the qubit decay through the port used to probe the readout resonator. The qubit is capacitively coupled to its transmission line at a rate $\Gamma_\mathrm{c}/2\pi= \SI{0.383}{kHz}$. This line is used to excite the qubit but also directly measure its emitted power, which contains the work extracted by the engine. The qubit presents a total relaxation time $T_1= 25.4 \pm 0.5~\si{\micro\second}$ and dephasing time $T_2$ varying from \SI{10}{\micro\second} to \SI{45}{\micro\second} over a few days. It is thus not Purcell-limited by its direct coupling to the transmission line (Purcell limit of \SI{416}{\micro s}). 

%%%%%%%%%%%%%%%%%%%%%%%%%%%%%%%%%%%%%%%%%%%

\section{Qubit state readout}
\label{app:RO}

\begin{figure}[h!]
    \centering
    \includegraphics[width=8.6cm]{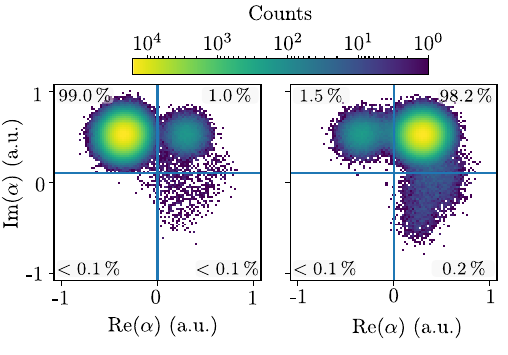}
    \caption{Qubit state readout: histograms of the readout quadratures when the qubit has been prepared in its thermal equilibrium state (close to its ground state $\ket{-z}$ in the left panel) or after a $\pi$-pulse from its thermal equilibrium state (close to its excited state $\ket{+z}$). }
    \label{fig:RO}
\end{figure}

 A nearly QND readout of $\hat{\sigma}_z$ is obtained in \SI{280}{ns} integration time with single-shot readout fidelity of $F = 1 - [P_{\vert -z \rangle}(+z) + P_{\vert +z \rangle}( -z) ]/2 =\SI{98.5}{\%}$ where $P_{\vert a \rangle}(b)$ is the probability to obtain the outcome $b$ when having prepared state $\ket{a}$. These probabilities are extracted from histograms of single-shot readouts, where the ground state $\ket{-z}$ is identified by having an outcome $\alpha$ in the top left part ($\Re{\alpha}<0$ and $\Im{\alpha}>0$) of the complex IQ plane while the excited state $\ket{+z}$ is identified as being in the top right part ($\Re{\alpha} \geq 0$ and $\Im{\alpha}>0$). Such readout histograms are shown in \cref{fig:RO} after having prepared the ground or the excited state. From the readout histograms of the thermal equilibrium state, we roughly estimate the qubit to be at an equilibrium temperature of \SI{54}{mK}, as the qubit has a thermal population around \SI{1}{\percent}.
Moreover, we have a probability $P_{\vert -z \rangle}(\overline{\pm z}) \sim \SI{0.1}{\%}$ and $P_{\vert +z \rangle}(\overline{\pm z}) \sim \SI{0.2}{\%}$ of not reading out the ground or excited state due to readout ionisation of the transmon out of its qubit subspace \cite{Sank2016, Lescanne2019, Shillito2022}. This $\SI{0.15}{\percent}$ error on average is not considered in the feedback protocol of our engine, where the transmon qubit will be assumed to only be in its ground state if $\Re{\alpha}<0$ or in its excited state if $\Re{\alpha} \geq 0$. We expect that it would lead to a small renormalization of the measured work.

As explained in the main text, this $\hat{\sigma}_z$ readout is transformed into a $\hat{\sigma}_x$ readout by applying a $-\pi/2$-pulse and then applying afterwards a $+\pi/2$ pulse.  
The $\pm \pi/2$-pulses used in the engine cycle and the $\pi$ pulses used in the qubit active reset have a hyperbolic secant temporal envelope whose width is $\sigma = \SI{8}{ns}$ and which is truncated at $4\sigma$. On previous cool downs, with the same sample, the qubit demonstrating a reduced $T_2$ of $\SI{7.5}{\mu s}$ had coherence-limited gate infidelities of \SI{0.16}{\percent} measured by randomized benchmarking \cite{Lledo2023}.

%%%%%%%%%%%%%%%%%%%%%%%%%%%%%%%%%%%%%%%%%%%
\section{AC-Stark shift and measurement-induced dephasing}
\label{app:dephasing}

\begin{figure}[h!]
    \centering
    \includegraphics[width=8.6cm]{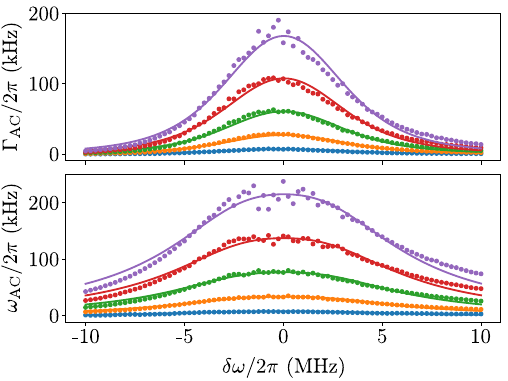}
    \caption{Measurement-induced dephasing $\Gamma_\mathrm{AC}$ and AC-Stark shift $\omega_\mathrm{AC}$ as a function of the detuning $\delta \omega$ of the driving tone compared to the readout resonator frequency. Experimental data in dots and fit using \cref{eq:AC_Stark} in solid line. The different colors correspond to different driving amplitudes, resulting in an average photon number at zero detuning of $3~ 10^{-3}$ in blue, $1.3~ 10^{-2}$ in orange, $3~ 10^{-2}$ in green, $5.3~ 10^{-2}$ in red, and $8.3~ 10^{-2}$ in purple. }
    \label{fig:meas-induced-dephasing}
\end{figure}

To characterize the couplings $\kappa$ and $\chi$ but also to calibrate the AC-Stark effect, we performed measurement-induced dephasing through a Ramsey-like pulse sequence~\cite{Campagne-Ibarcq2013}.
The induced frequency shift $\omega_\mathrm{AC}$ and dephasing $\Gamma_{AC}$ are given by
\begin{align}
    \label{eq:AC_Stark}
    \Gamma_\mathrm{AC} = \chi \Im{\alpha_{-z}^* \alpha_{+z}}
    \notag \\
    \omega_\mathrm{AC} = \chi \Re{\alpha_{-z}^* \alpha_{+z}}
\end{align}
where $\alpha_{-z} = \epsilon /(\kappa/2 -i(\delta\omega -\chi/2) )$ and $\alpha_{+z} = \epsilon /(\kappa/2 -i(\delta\omega +\chi/2) )$ are the readout resonator coherent steady-state when the qubit is in state $\ket{-z}$ or $\ket{+z}$ respectively. The driving amplitude $\epsilon$ and detuning $\delta \omega = \omega_r -\omega_d$ between the driving frequency $\omega_d$ and the readout resonator frequency $\omega_r$. They are displayed in \cref{fig:meas-induced-dephasing} for different input amplitudes. The fit using \cref{{eq:AC_Stark}} allows us to calibrate the attenuation factor between the room temperature amplitude and $\epsilon$ but also to extract the rates $\chi/2\pi = \SI{3.3}{MHz}$ and $\kappa/2\pi = \SI{9}{MHz}$. 
During the engine calibration sequence to measure $P_\mathrm{in}$, the qubit is shifted by $\omega_\mathrm{AC}/2\pi = \SI{1}{MHz}$ (and $\Gamma_\mathrm{AC}/2\pi = \SI{0.14}{MHz}$) with the calibrated driving amplitude at a detuning of $\delta\omega/2\pi = \SI{-10}{MHz}$. 

\section{Calibration of the measured power}
\label{app:calib_P}

\begin{figure}[h!]
    \centering
    \includegraphics[width=\columnwidth]{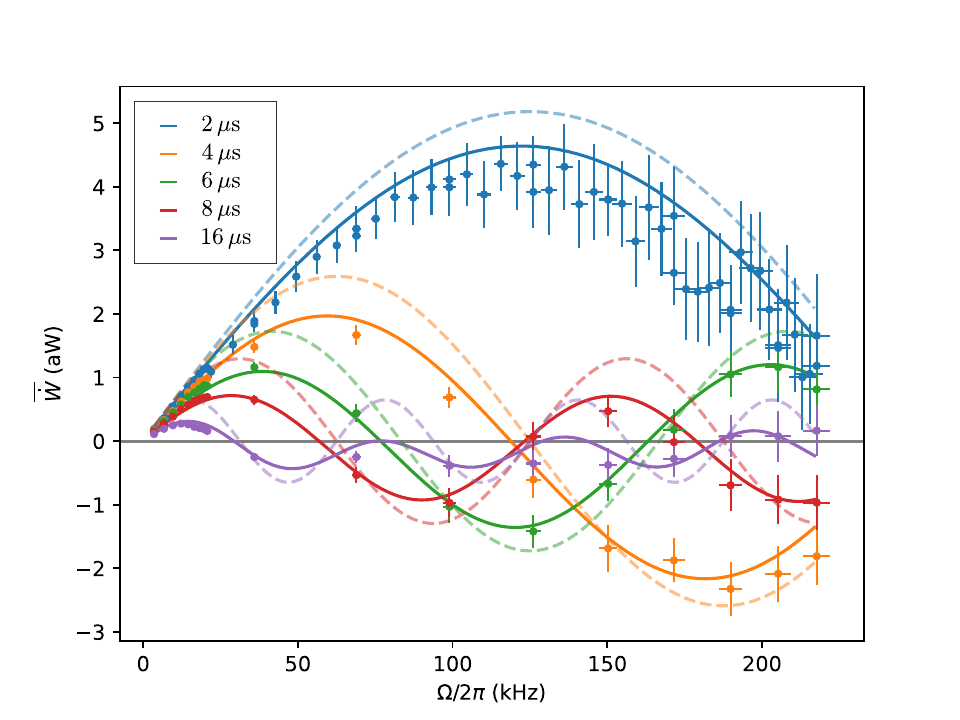}
    \caption{Dots: Measured extracted work flow $\overline{W}/t_R$ in aW as a function of Rabi rate $\Omega$. The different colors correspond to different durations $t_\mathrm{R}$ (see legend). Dashed lines: ideal case of infinitely long lived qubit (as in Fig.~\ref{fig:work_theta}). Solid lines: numerical predictions already represented in Fig.~\ref{fig:work_theta}b.  }
    \label{fig:work_real_units}
\end{figure}

\begin{figure*}[t!]
    \centering
    \includegraphics[width=17.2cm]{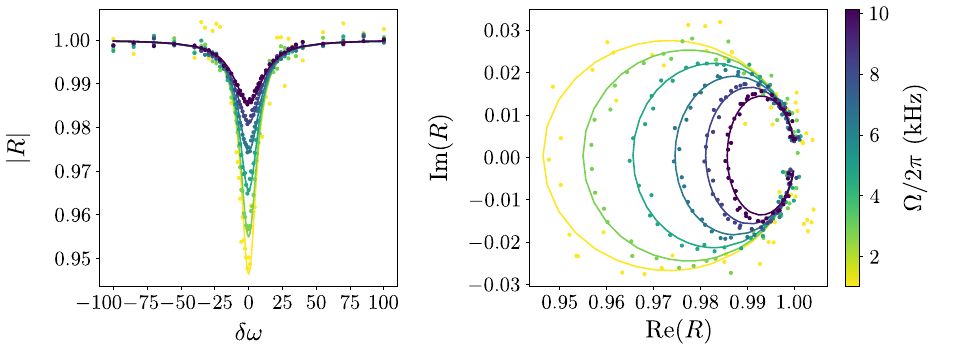}
    \caption{Amplitude (a) and polar representation (b) of the qubit reflection coefficient $R$ as a function of detuning $\delta\omega = \omega_\mathrm{q} - \omega_\mathrm{d}$ between the drive $\omega_\mathrm{d}$ and the qubit frequency $\omega_\mathrm{q}$ for various drive amplitudes, i.e. various Rabi rate $\Omega$. Dots: measured and solid lines: fits using ~\cref{eq:reflection_coefficient}.}
    \label{fig:fluo_reflection}
\end{figure*}
With our heterodyne measurement setup, the field that we are measuring is not $\hat{b}_\text{out}$ but the field $\hat{B}_\text{out}$ after the output amplification line following
\begin{equation}
\hat{B}_\mathrm{out} = \sqrt{G_\text{meas}} \hat{b}_\text{out} + \sqrt{G_\text{meas}-1} \hat{i}^\dagger_\mathrm{out}
\end{equation}
where $G_\text{meas}$ is the total gain of the amplifiers involved in the measurement setup and $\hat{i}_\mathrm{out}$ is an effective idler mode, that captures the addition of noise of phase-preserving amplification and verifies $\expval{\hat{i}_\text{out}}=0$.
The output power is thus given by
\begin{equation}
    \left|\expval{\hat{B}_\mathrm{out}}\right|^2= G_\text{meas} \left( |\beta_\mathrm{in}|^2 + \frac{\Omega}{2}\expval{\sigma_x}+ \Gamma_\mathrm{c} \frac{ \expval{\sigma_x}^2}{4}\right)
\end{equation}
where $\expval{\hat{b}_\mathrm{in}}=\beta_\mathrm{in}$. When using the AC Stark effect to detune the qubit: 
\begin{equation}\left|\expval{\hat{B}_\mathrm{in}}\right|^2= G_\text{meas}  |\beta_\mathrm{in}|^2.
\end{equation}

From the measured quadratures $\expval{\hat{B}_\mathrm{in/out} (t)}$ and calibrations of the gain $G_\text{meas}$ for the different sets of measurements, we compute the excess power 
\begin{equation}
   \frac{P_\mathrm{out}(t) - P_\mathrm{in}(t) }{\hbar \omega_\mathrm{q} } = \frac{\left\vert \expval{ \hat{B}_\mathrm{out} (t)} \right\vert^2 - \left\vert \expval{ \hat{B}_\mathrm{in} (t)} \right\vert^2}{G_\text{meas}}.
\end{equation}
We also consider drifts of the gain of the detection setup during the measurement of the engine. To mitigate this effect, during the pulse sequence used to determine the average work extraction of the engine, we calibrate the gain $G_\text{meas} = 4 \Gamma_c \left(\frac{  \left\vert \expval{\hat{B}_\mathrm{in}(\alpha)}\right\vert  }{\Omega(\alpha)}\right)^2$ by measuring the output field amplitude $\expval{\hat{B}_\mathrm{in}}$ as a function of the set digital amplitude $\alpha$ . We assume the Rabi frequency $\Omega(\alpha)$ and $\expval{\hat{b}_\mathrm{in}}$ to be constant in time. Moreover, to take into account the non-linearity of the room temperature mixers, both $\expval{B_\mathrm{in}}(\alpha)$ and $\Omega(\alpha)$ are fitted with order 2 polynomials of $\alpha$.
%%%%%%%%%%%%%%%%%%%%%%%%%%%%%%%%%%%%%%%%%%%

It is interesting to plot the data measured in Fig.~\ref{fig:work_theta} as a function of Rabi frequency $\Omega$ and in actual power units. In the figure~\ref{fig:work_real_units}, one can appreciate that the power we measure are of the order of a few aW only.

\section{Calibration of the coupling rate $\Gamma_\mathrm{c}$}
\label{app:spectro_fluo}

The rate $\Gamma_\mathrm{c}$ characterizes the coupling rate of the qubit with the transmission line. We determine it by measuring the reflection coefficient $R = \frac{\expval{\hat{b}_\text{out}}}{\expval{\hat{b}_\text{in}}}$ of the qubit~\cite{Stevens2022}
\begin{equation}
R(\delta) =  1 - (p_{\ket{-z}}^{\text{th}}-p_{\ket{+z}}^{\text{th}}) \frac{\Gamma_\mathrm{c} \Gamma_1 (\Gamma_2 - i \delta)}{\left[\Gamma_1 (\Gamma_2^2 + \delta^2)+\Gamma_2\Omega^2\right]} \ , \label{eq:reflection_coefficient}
\end{equation}
where $\Gamma_1 = 1/T_1$, $\Gamma_2 = 1/T_2$, $\delta\omega = \omega_\mathrm{q} - \omega_\mathrm{d}$. The thermal population $p_{\ket{+z}}^\text{th} = \SI{1}{\percent}$ is suppressed using active reset heralding. 
The coherence times $T_1$ and $T_2$ are measured using standard relaxation and Ramsey interferometry techniques. While the input and output lines introduce scaling factors between the fields $\expval{\hat{b}_\text{in}}$, $\expval{\hat{b}_\text{out}}$ and the fields that we actually send and measure, it can be taken into account by measuring $R$ for large enough detuning $\delta$. This leaves $\Omega$ and $\Gamma_\mathrm{c}$ as the only fit parameters in \cref{eq:reflection_coefficient}. We show $R(\delta)$ for various Rabi frequencies $\Omega$ in \cref{fig:fluo_reflection}, from which we extract $\Gamma_\mathrm{c}/2\pi = \SI{0.383}{\kilo\hertz}$.
Due to the fluctuating qubit parameters over time, the averaging on this reflection measurement is interleaved with recalibration measurements of the qubit parameters. The reflection measurement has been performed in another cool-down than the experiments discussed in the main text.

\section{Qubit Purcell decay through the readout resonator}
Even though there is a Purcell filter, the qubit is also coupled to the second transmission line via its Purcell decay $\Gamma_\mathrm{P}$ through the readout resonator that we estimate here. By comparing the Rabi rates $\Omega_1 = 2 \sqrt{\Gamma_\mathrm{c}} b_\mathrm{in, 1}$ and $\Omega_2 = 2 \sqrt{\Gamma_\mathrm{P}} b_\mathrm{in, 2}$ from driving the qubit through the "qubit port" or the "readout port" respectively, and considering the different attenuations of the two input lines, we extract a Purcell decay through the readout resonator of $\Gamma_\mathrm{P}/2\pi \simeq \SI{10}{Hz} \ll \Gamma_\mathrm{c}$. The qubit total decay rate $T_1^{-1} = 2\pi \cdot \SI{6.4}{kHz} \gg \Gamma_\mathrm{c} \gg \Gamma_\mathrm{P}$ is thus not limited by its couplings to the microwave transmission lines. Moreover, due to the diplexer (see \cref{fig:wiring}), the output line only amplifies signals coming from $\Gamma_\mathrm{c}$ and not the one from $\Gamma_\mathrm{P}$.

\section{Extracted work as a function of cycle number}
\label{app:PvsNc}

\begin{figure}[h!]
    \centering
    \includegraphics[width=8.6cm]{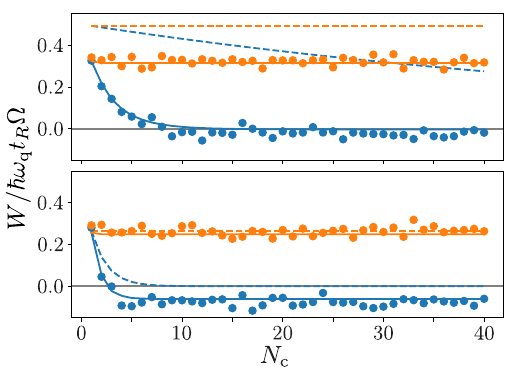}
    \caption{ Dots: normalized extracted work $W/\hbar \omega_\mathrm{q} t_R \Omega$ of the engine in orange and of the open-loop configuration (blue) as a function of cycle number $N_c$ for $t_R = \SI{8}{\mu s}$ and $\Omega/2\pi = \SI{3.4}{kHz}$ (top) or $\Omega/2\pi = \SI{20.2}{kHz}$ (bottom). Dash lines are the analytical extracted work $W^{(N_c)} = \frac{\hbar \omega_\mathrm{q} \theta}{2} \cos^{N_c}(\theta)$ while solid lines are simulations. }
    \label{fig:P_vs_Nc}
\end{figure}

We measure the extracted work per cycle $W(N_c)$ as a function of cycle number $N_c$. In the engine configuration, the extracted work is as intended, independent of the cycle number (\cref{fig:P_vs_Nc}). However, the qubit's imperfections modify the amount of extractable work for a given $t_R$ and $\Omega$. In the open-loop configuration, we observe a much faster decay of the extracted work than the expected decrease $W(N_c) = \frac{\hbar \omega_\mathrm{q} \theta}{2} \cos^{N_c}(\theta)$ for a dissipationless qubit. In both the engine and open-loop configuration, the simulations fit well with the extracted work.

\section{Average per cycle extracted work in open-loop configuration}
\label{app:feedbackoff}

\begin{figure}[h!]
    \centering
    \includegraphics[width=8.6cm]{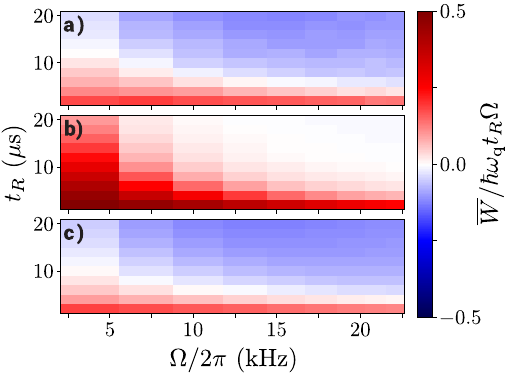}
    \caption{ Normalized extracted work $\overline{W}/\hbar\omega_\mathrm{q} t_R \Omega$ averaged over the first 40 cycles in case of open-loop configuration, measured in a), analytical in b) and simulated in c), as a function of duration $t_R$ and amplitude $\Omega$. 
    }
    \label{fig:feedback_OFF_T1_effect}
\end{figure}

Due to finite dissipation, the Zeno limit $\theta \rightarrow 0$, where work can be extracted even in the open-loop configuration, can still be approached but is less pronounced (see \cref{fig:feedback_OFF_T1_effect}). Without dissipation, the extracted work tends to zero in the steady state (for a large cycle number $N_c \rightarrow +\infty$) except for the Zeno limit $\theta \rightarrow 0$. When considering dissipation, a finite non-zero work can still be extracted in the steady state. This work can be positive or negative depending on the Rabi parameters $\Omega$ and $t_R$ for a given amount of dissipation.

\section{Characterization of the amplification process}
\label{app:ampli}

\subsection{Time and phase dependence}

To characterize the properties of the amplification process, we consider the case of an input pulse of arbitrary phase $\phi$, such that $\expval{\hat b_\text{in}} = \frac{\Omega}{2\sqrt{\Gamma_c}}e^{i\phi}$ (while we assume the measurement and feedback step are still chosen to prepare the state $\ket{+_x}$).
One obtains: 
\begin{align}
    \frac{P_\text{out}}{\hbar\omega_\text{q}}&=\left|\expval{\hat b_\text{out}}\right|^2\notag\\
&= \frac{P_\text{in}}{\hbar\omega_\text{q}} + \frac{\Omega}{2}\left(\cos\phi \expval{\sigma_x(t)}-\sin\phi\expval{\sigma_y(t)}\right)\notag\\
&+\Gamma_c\left|\expval{\sigma_-(t)}\right|^2.
\end{align}
In the same limit $\Omega\gg \Gamma_c$ as in the main text, we neglect the last term and  obtain:
$$
G(t) -1 = \frac{2\Gamma_c}{\Omega} \left(\cos\phi \expval{\sigma_x(t)}-\sin\phi\expval{\sigma_y(t)}\right),
$$
which depends explicitly of $\phi$. We further analyze the phase dependence by focusing on the model with fixed qubit coherent time $T_2$, where one finds that 
$$\cos\phi \expval{\sigma_x(t)}-\sin\phi\expval{\sigma_y(t)} = \cos\phi \cos\left(\Omega t\right)e^{-t/T_2},$$
such that 
$$
G(t) -1 = \frac{2\Gamma_c}{\Omega} \cos\phi \cos\left(\Omega t\right)e^{-t/T_2}.
$$

The gain is therefore phase-dependent, for a fixed measurement basis (or requires to tune the measurement basis and the phase of the pulse to amplify). 

The gain is also time-dependent over the duration of the input pulse. In the following, we will consider the case of short duration pulses $t_R \ll 1/\Omega, T_2$, so that $\cos\left(\Omega t\right)e^{-t/T_2} \simeq 1$ and $\phi=0$, where the excess gain is given by $G-1 \simeq \frac{2\Gamma_c}{\Omega}$.

\subsection{Input power dependence}
The gain $G = 1 + \frac{2\Gamma_c}{\Omega}$ is nonlinear as its value depends on the input power through $\Omega = 2 \sqrt{\Gamma_\mathrm{c}} \abs{\expval{\hat{b}_\mathrm{in}}} = 2 \sqrt{\Gamma_\mathrm{c}\frac{P_\mathrm{in}}{\hbar\omega_\mathrm{q}} }$.
The gain of the amplification will saturate to 1 for large input power.

\subsection{Output field variance}

\begin{figure}[h!]
    \centering
    \includegraphics[width=8.6cm]{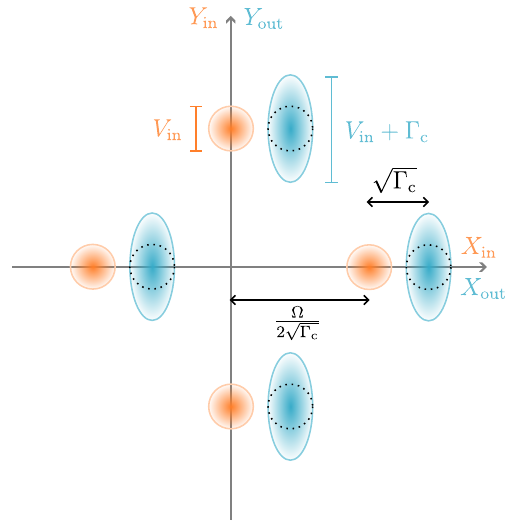}
    \caption{ Schematic representation of the input and output signals in the quadrature phase space. In orange are shown the Wigner functions of  four possible coherent states of the input field of amplitude $\expval{\hat{b}_\mathrm{in}} = \epsilon \frac{\Omega}{2\sqrt{\Gamma_\mathrm{c}}}$, with $\epsilon=1,i,-1$ or $-i$, and with variance $V_\mathrm{in}$, have been represented. In blue are shown the corresponding Wigner functions of the output signals, displaced towards quadrature +X, and with a variance increased by $\Gamma_\mathrm{c}$ along the Y quadrature.
    }
    \label{fig:IQ_schematic}
\end{figure}

We introduce the two quadratures of the output mode:
\begin{eqnarray}
X_\text{out}&=b_\text{out}+b_\text{out}^\dagger =b_\text{in}+b_\text{in}^\dagger +\sqrt{\Gamma_\mathrm{c}}\sigma_x\nonumber\\
Y_\text{out}&=i(b_\text{out}-b_\text{out}^\dagger) =i(b_\text{in}-b_\text{in}^\dagger) +\sqrt{\Gamma_\mathrm{c}}\sigma_y.
\end{eqnarray}
Using that the input mode is not correlated with the atom state, we find their variances:
\begin{eqnarray}
\langle X_\text{out}^2\rangle-\langle X_\text{out}\rangle^2 &=\langle X_\text{in}^2\rangle-\langle X_\text{in}\rangle^2+\Gamma_\mathrm{c}(1-\expval{\sigma_x(t)})\nonumber\\
\langle Y_\text{out}^2\rangle-\langle Y_\text{out}\rangle^2&=\langle Y_\text{in}^2\rangle-\langle Y_\text{in}\rangle^2+\Gamma_\mathrm{c}(1-\expval{\sigma_y(t)}).\nonumber\\
\end{eqnarray}
In the limit of a short pulse $t_R \ll 1/\Omega$, such that $\expval{\sigma_x(t)}\simeq 1$, $\expval{\sigma_y(t)}\simeq 0$, the $X_\text{out}$ quadrature which contains the amplified signal does not receive any additional noise, which is exclusively added to the $Y_\text{out}$ quadrature (\cref{fig:IQ_schematic}).

\section{Qubit parameters stability}
\label{app:stability}

\begin{figure}[h!]
    \centering
    \includegraphics[width=8.6cm]{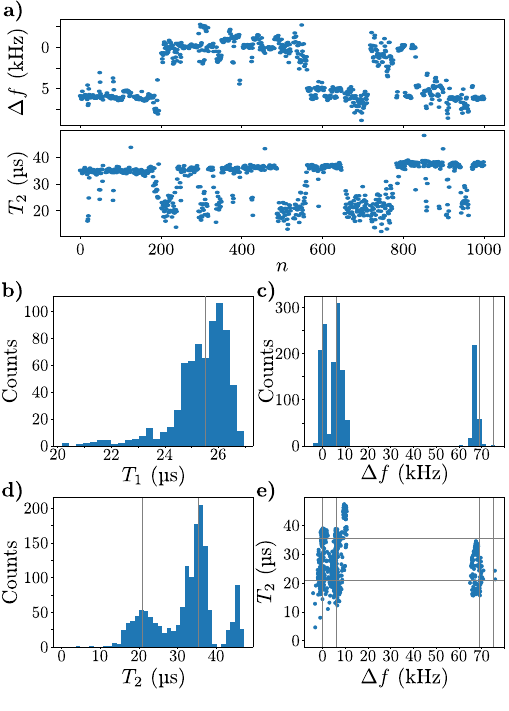}
    \caption{Qubit parameters instability. In a) qubit detuning $\Delta f$ and coherence time $T_2$ as function of measurement number $n$ where each measurement takes approximately \SI{1}{min}.
    Histograms of $T_1$ in b), $\Delta f$ in c) and $T_2$ in d) of compiled measurements from different datasets. The different sets are spaced several days up to weeks apart. In e), scatter plot of $T_2$ as a function of $\Delta f$ which are both extracted from the same Ramsey measurements.  }
    \label{fig:stabilities}
\end{figure}

We have observed that the qubit parameters were not stable. Through repeated relaxation and Ramsey interferometry measurements over time, we extract the histograms of the qubit parameters $\omega_\mathrm{q}$, $T_1$ and $T_2$.  

Different distributions could be noticed in the qubit $\omega_\mathrm{q}$ and $T_2$, that we attribute to the coupling to parasitic two-level systems~\cite{Muller_2019}. In \cref{fig:stabilities} are shown the qubit parameters extracted from compiled data of different sets of relaxation and Ramsey measurements taken at a rate of approximately \SI{1}{\per\minute}. The different sets are spaced several days up to weeks apart and were performed in the same cool-down as the experiments discussed in the main text.
One of this dataset, displayed in \cref{fig:stabilities}.a, present the qubit coherence $T_2$ and qubit detuning $\delta f$ as function of measurement number, each measurement taking about \SI{1}{min}. 

\section{Simulations}
\label{app:simulations}
The simulations are implemented with the QuantumOptics.jl library in Julia simulating the master equation evolution of the qubit under the different drives and feedback. In the simulations, only the Hilbert space of the qubit is considered. The extracted work is computed from the qubit density matrix as $\frac{P(t)}{\hbar \omega_\mathrm{q}} = \frac{\Omega}{2}\expval{\hat{\sigma}_x(t)}$ during the Rabi driving pulse. The whole time evolution of the pulse sequence of $N_c$ cycles is computed as described in \cref{fig:principle}.c). 

For the dispersive readout pulse, it is simplified and decomposed into three steps: first, a free evolution during $t_\mathrm{int}/2$, then an instantaneous projection over the states $\ket{-z}$ or $\ket{+z}$ followed again by a free evolution during $t_\mathrm{int}/2$. The simulations also incorporate imperfections such as an assignation error of \SI{0.4}{\percent} in the readout and the qubit thermalization to a \SI{1}{\%} population.

%%%%%%%%%%%%%%%%%%%%%%%%%%%%%%%%%%%%%%%%%%%

%

\end{document}